\documentclass[prd,twocolumn,twoside,preprintnumbers,superscriptaddress,nofootinbib]{revtex4}

\usepackage{amsmath}
\usepackage{graphicx}
\usepackage{multirow}
\usepackage{color}
\usepackage{cool}
\usepackage{hyperref}
\usepackage{setspace}
\usepackage{enumitem}
\usepackage{amssymb}
\usepackage[normalem]{ulem}

\def\be{\begin{equation}}
\def\ee{\end{equation}}
\def\beq{\begin{equation}\begin{aligned}}
\def\eeq{\end{aligned}\end{equation}}

\newcommand{\Eq}[1]{Eq.~\eqref{#1}}

\def\gev{\, {\rm GeV}}

\begin{document}
%%%%%%%%%%%%%%%%%%%%%%%%%%%%%%%%%

\preprint{\vbox{\hbox{SCIPP 17/02}}}

\title{When the Universe Expands Too Fast: Relentless Dark Matter}

\author{Francesco~D'Eramo}
\email{fderamo@ucsc.edu}
\affiliation{Department of Physics, 1156 High St., University of California Santa Cruz, Santa Cruz, CA 95064, USA}
\affiliation{Santa Cruz Institute for Particle Physics, 1156 High St., Santa Cruz, CA 95064, USA}

\author{Nicolas~Fernandez}
\email{nfernan2@ucsc.edu}
\affiliation{Department of Physics, 1156 High St., University of California Santa Cruz, Santa Cruz, CA 95064, USA}
\affiliation{Santa Cruz Institute for Particle Physics, 1156 High St., Santa Cruz, CA 95064, USA}

\author{Stefano~Profumo}
\email{profumo@ucsc.edu}
\affiliation{Department of Physics, 1156 High St., University of California Santa Cruz, Santa Cruz, CA 95064, USA}
\affiliation{Santa Cruz Institute for Particle Physics, 1156 High St., Santa Cruz, CA 95064, USA}

\date{March 13, 2017}
%%%%%%%%%%%%%%%%%%%%%%%%%%%%%%%%%%
%%%%%%%%%%%%%%%%%%%%%%%%%%%%%%%%%%
\begin{abstract}

We consider a modification to the standard cosmological history consisting of introducing a new species $\phi$ whose energy density red-shifts with the scale factor $a$ like $\rho_\phi \propto a^{-(4+n)}$. For $n>0$, such a red-shift is faster than radiation, hence the new species dominates the energy budget of the universe at early times while it is completely negligible at late times. If equality with the radiation energy density is achieved at low enough temperatures, dark matter can be produced as a thermal relic during the new cosmological phase. Dark matter freeze-out then occurs at higher temperatures compared to the standard case, implying that reproducing the observed abundance requires significantly larger annihilation rates. Here, we point out a completely new phenomenon, which we refer to as {\em relentless} dark matter: for large enough $n$, unlike the standard case where annihilation ends shortly after the departure from thermal equilibrium, dark matter particles keep annihilating long after leaving chemical equilibrium, with a significant depletion of the final relic abundance. Relentless annihilation occurs for $n \geq 2$ and $n \geq 4$ for $s$-wave and $p$-wave annihilation, respectively, and it thus occurs in well motivated scenarios such as a quintessence with a kination phase. We discuss a few microscopic realizations for the new cosmological component and highlight the phenomenological consequences of our calculations for dark matter searches.%We call candidates with this peculiar behavior relentless dark matter, because of their continuos attempt to get back to thermal equilibrium. 
%We provide microscopic realizations of such a new fluid, and highlight the phenomenological consequences of our calculations for dark matter searches. 

\end{abstract}

\maketitle

\section{Introduction}

Decades after the first observational evidences, the origin and composition of the dark matter (DM) is still among the most urgent open questions in particle physics~\cite{Jungman:1995df,Bertone:2004pz,Feng:2010gw}. Weakly Interacting Massive Particles (WIMPs) are motivated particle candidates for DM, with a thermal relic abundance naturally close to the observed cosmological abundance of DM. A standard calculation~\cite{Lee:1977ua,Scherrer:1985zt,Srednicki:1988ce} shows that this thermal relic WIMP abundance scales as the inverse annihilation cross section, and is mildly dependent on the particle mass. The observed DM abundance is reproduced for
\be
\langle \sigma_{\rm th} v_{\rm rel} \rangle \simeq 3 \times 10^{-26} \, \text{cm}^3 \, \text{sec}^{-1} \ ,
\label{eq:sigmavrelthermal}
\ee
where the brackets denote a thermal average and $v_{\rm rel}$ is the M\o{}ller velocity (for details see Ref.~\cite{Gondolo:1990dk}). The cross section needed for a thermal relic is thus that typical of weak interactions. This phenomenal coincidence, combined with the expectation of new degrees of freedom at the weak scale for independent reasons such as the hierarchy problem, is referred to as the ``WIMP miracle''. 

The numerical value in \Eq{eq:sigmavrelthermal} has been an important benchmark for WIMP searches. It is worth keeping in mind that it relies on a crucial assumption: at the time of DM genesis, the energy budget of the universe was dominated by its radiation content. We know from Big Bang Nucleosynthesis (BBN) that this is definitely the case at temperatures around and below $T_{\rm BBN} \simeq \text{few} \; {\rm MeV}$~\cite{Kawasaki:2000en,Ichikawa:2005vw}. However, we have no direct information about the energy budget of the universe at higher temperatures. The WIMP DM thermal relic abundance may differ by orders of magnitude if deviations from a standard cosmological history are considered~\cite{McDonald:1989jd,Kamionkowski:1990ni,Chung:1998rq,Giudice:2000ex,Moroi:1999zb,Allahverdi:2002nb,Allahverdi:2002pu,Acharya:2009zt,Monteux:2015qqa,Co:2015pka,Davoudiasl:2015vba,Co:2016vsi,Co:2016fln}.

In this work we consider DM genesis for a broad class of alternative cosmological histories. We assume the presence of another species $\phi$, whose energy density red-shifts with the scale factor $a$ as follows
\be
\rho_\phi \propto a^{-(4+n)} \ , \qquad \qquad n > 0 \ .
\label{eq:phiScaling}
\ee
The standard case of radiation follows the behavior above for $n=0$. Here, we always consider $n > 0$, which implies that the $\phi$ energy density dominates over radiation at early enough times. The equality between the energy density of $\phi$ and radiation must happen at a temperature $T_r \gtrsim T_{\rm BBN}$ (we will be more quantitative about this point in  Sec.~\ref{sec:BBN}). If such an equality is achieved after the time of DM production, the standard relic calculation is significantly affected, as is the thermal relic abundance of the DM. We survey the options for DM genesis when the universe is dominated by a fluid red-shifting as in \Eq{eq:phiScaling} in Sec.~\ref{sec:FO}.

The two-dimensional parameter space $(T_r, n)$ fully describes the possible cosmological backgrounds in our setup. The two parameters cannot be arbitrary, since for low enough $T_r$ we must ensure not to spoil the success of BBN. This set of cosmological backgrounds are described in Sec.~\ref{sec:background}, where we provide an expression for the Hubble parameter as a function of the radiation bath temperature $T$. For each temperature value $T > T_r$, the Hubble parameter is always larger than what it would be for a standard cosmological history. For this reason, the universe expands {\em faster} than in the standard case when dominated by $\phi$.

A significant energy density of $\phi$ around the time of BBN mimics the role of additional neutrino species. Light element abundances put bounds on $N_\nu$~\cite{Cyburt:2015mya}, which can be used to exclude part of the $(T_r, n)$ plane. We discuss these bounds in Sec.~\ref{sec:BBN}. Interestingly, the energy density of $\phi$ is completely subdominant at the time of the decoupling of the Cosmic Microwave Background (CMB). The number of effective neutrinos at $T_{\rm CMB} \simeq 1 \, {\rm eV}$ is also constrained~\cite{Ade:2015xua}, but our framework does not predict any deviation from the SM value. %However, we can still detect a deviation from the SM value at BBN.

This work focuses on freeze-out DM production. We assume the DM particles to achieve thermal equilibrium with the primordial plasma at high temperature, and decouple once the temperature drops below its mass. The faster expansion rate raises however an important question: does the DM {\em ever}= thermalize? This is inspected in App.~\ref{app:DMthermalization}, where we quantify the conditions we need in order to have the DM in thermal equilibrium at early times. The answer to this question sets the stage for the DM relic density calculations in Sec.~\ref{sec:FO}. These calculations are performed by parameterizing the new cosmological phase by $(n, T_r)$, without specifying the microscopic origin of the new species $\phi$. At large enough $n$, we find a very peculiar behavior for the DM number density evolution, that had never been recognized before: The different Hubble scaling with the temperature allows significant DM annihilations long after the decoupling from the thermal bath. For a DM annihilating through an $s$-($p$-)wave process, this happens for $n \geq 2 (4)$. Remarkably, the red-shift with $n = 2$ arises from motivated theories of quintessence attempting to explain the current acceleration of our universe~\cite{Caldwell:1997ii,Sahni:1999gb}. We call relic particles freezing-out during this phase {\it relentless} dark matter, due to their obstinate struggle to get back to thermal equilibrium. This behavior, which we find in our numerical results shown in Figs.~\ref{fig:FreezeOut} and \ref{fig:FreezeOut_p}, is easily understood with the semi-analytical results given in App.~\ref{app:analytical}. Notably, the phenomenon of relentless dark matter leads to {\em significant numerical differences} in the calculation of the thermal relic density for example in the case of kination-domination phases from previous studies (see e.g. \cite{Salati:2002md,Profumo:2003hq}).

The faster expansion rate implies an earlier freeze-out. Since we are dealing with cold relic, reproducing the observed DM density requires couplings significantly larger than in the standard case. This opens up the possibility of having cross section substantially larger than the thermal value in \Eq{eq:sigmavrelthermal}, in contrast with the case of an early matter-dominated epoch providing dilution where smaller values of the cross section are required~\cite{McDonald:1989jd,Kamionkowski:1990ni,Chung:1998rq,Giudice:2000ex}, and consequently weaker signals in DM searches. We quantify how much annihilation cross sections can be enhanced in Figs.~\ref{fig:Cross_Section_s} and \ref{fig:Cross_Section_p}. 

Finally, we address the question of the origin of the new cosmological component $\phi$ in Sec.~\ref{sec:EFC}, where we provide one explicit example of a microscopic theory leading to the behavior in \Eq{eq:phiScaling}. We summarize our results in Sec.~\ref{sec:Conclusions}, where we also discuss future work addressing the implications of our analysis for dark matter searches.

\section{A faster expansion}
\label{sec:background}

The expansion rate of the universe, quantified by the Hubble parameter $H$, is controlled by its energy density through Friedmann's equations. We consider cosmological histories where two different species populate the early universe, radiation and $\phi$, with a total energy density $\rho = \rho_{\rm rad} + \rho_\phi$.

The contribution from radiation, the only one present for a standard cosmological history, can be expressed in terms of its temperature as follows 
\be
\rho_{\rm rad}(T) = \frac{\pi^2}{30} g_*(T) T^4 \ ,
\label{eq:rhorad}
\ee
where $g_*(T)$ is the number of effective relativistic degrees of freedom. We find it useful to express $\rho_\phi$ as a function of the radiation temperature $T$. All we know is its red-shift behavior given in Eq.~\eqref{eq:phiScaling}, hence we need to connect $a$ with $T$. This is achieved by assuming and imposing entropy conservation in a comoving volume $S = s a^3 = {\rm const}$, where the entropy density reads
\be
s(T) =  \frac{2 \pi^2}{45} g_{*s}(T) T^3 \ .
\ee
Here, $g_{*s}$ is the effective relativistic degrees of freedom contributing to the entropy density. Entropy conservation ensures $g_{*s}(T)^{1/3} T a = {\rm const}$, and the scaling in Eq.~\eqref{eq:phiScaling} can be re-expressed as follows
\be
\rho_\phi(T) = \rho_\phi(T_r)  \left(\frac{g_{*s}(T)}{g_{*s}(T_r)}\right)^{(4+n)/3} \left(\frac{T}{T_r}\right)^{(4+n)} \ .
\ee
Here, $T_r$ is some reference temperature set by the boundary conditions. We choose $T_r$ as the temperature where the two fluids have equal energy densities. The full energy density at any temperature reads
\be
\begin{split}
& \rho(T) = \rho_{\rm rad}(T) + \rho_\phi(T) = \\ & 
\rho_{\rm rad}(T) \left[1 + \frac{g_*(T_r)}{g_*(T)} \left( \frac{g_{*s}(T)}{g_{*s}(T_r)} \right)^{(4+n)/3} \left(\frac{T}{T_r}\right)^{n} \right] \ ,
\label{eq:rhototal}
\end{split}
\ee 
where we find it convenient to factor out the energy density of the radiation bath as given in \Eq{eq:rhorad}. From this expression it is manifest that the energy budget of the universe is dominated by $\phi$ for temperatures $T \gtrsim T_r$. 

With \Eq{eq:rhototal} in hand, we can evaluate the Hubble parameter as a function of the temperature
\be
H = \frac{\sqrt{\rho}}{\sqrt{3} \, M_{\rm Pl}}  \ ,
\label{eq:HubbleRate}
\ee
where the reduced Planck mass is $M_{\rm Pl} = (8 \pi G)^{-1/2} = 2.4 \times 10^{18} \, {\rm GeV}$. At temperatures larger than $T_r$, and setting for simplicity $g_*(T) = \overline{g}_* = {\rm const}$, the Hubble rate approximately is
\be
H(T) \simeq \frac{\pi \, \overline{g}^{1/2}_*}{3 \sqrt{10}} \frac{T^2}{M_{\rm Pl}} \left(\frac{T}{T_r}\right)^{n/2} \ ,  \qquad
(T \gg T_r) \ ,
\label{eq:HubbleEarlyTimes}
\ee
The full standard model (SM) degrees of freedom would lead $\overline{g}_* = g_{*{\rm SM}} =106.75$. The expression in \Eq{eq:HubbleEarlyTimes} is the Hubble rate at the time of DM genesis for the cosmological histories considered in this work. This result manifestly shows how the expansion rate at a given temperature $T$ is always larger than the correspondent value for a standard cosmological history. In our numerical analysis we use the complete expression for $H$, including the full temperature dependence of $g_*(T)$. 

\section{BBN Constraints}
\label{sec:BBN}

The successful predictions of light element abundances give us a quantitative test of the energy content of the universe when it was few seconds old. Before we consider freeze-out in the cosmological background described in Sec.~\ref{sec:background}, we have to ensure that we do not spoil this remarkable agreement between theoretical predictions and observations.

A potential issue with BBN arises if $T_r$ is not too far away from the ${\rm MeV}$ scale, where light elements begin to form. If this is the case, the universe expands faster than the usual case around the BBN time, and the theoretical prediction for BBN abundances may be altered. 

We parameterize the effect of the field $\phi$ by an effective number of relativistic degrees of freedom
\be
\rho(T) = \frac{\pi^2}{30} g^{\rm eff}_*(T) T^4 \ ,
\label{eq:rhototal2}
\ee
where we define
\be
g^{\rm eff}_*(T) = g_*(T) + \Delta g_*^\phi(T)  \ .
\ee
Here, $g_*(T)$ is the standard contribution from radiation, whereas $\Delta g_*^\phi(T)$ accounts for the energy density of $\phi$. The expression for the total energy density in Eq.~\eqref{eq:rhototal} define unambiguously the latter. A historical and widely used way to parameterize this effect is to describe the presence of $\phi$ as the the number of effective neutrinos. Within this convention, the total number of relativistic degrees of freedom appearing in \Eq{eq:rhototal2} reads
\be
g^{\rm eff}_*(T) = 2 + \frac{7}{8} \times 4 + \frac{7}{8} \times 2 \times N_\nu \ .
\label{eq:geffdef}
\ee
Here, we account for photons and positrons as well as neutrinos. In the absence of physics beyond the SM, the number of neutrino flavors at temperatures $T > 1 \, {\rm MeV}$ is $N_\nu^{({\rm SM})} = 3$.~\footnote{At lower temperatures, neutrinos decouple from the thermal bath, and after $e^+ e^-$ pair annihilations their temperature is lower than the photons, $T_\nu = (4/11)^{1/3} T_\gamma$. Furthermore, corrections due to non-instantaneous neutrino decoupling lead to a SM effective number of neutrino light flavors $N_{\rm eff}^{({\rm SM})} = 3.04$~\cite{Dodelson:1992km}.} By a comparison between the last two expressions, we compute $\Delta N_\nu  \equiv N_\nu - N_\nu^{({\rm SM})} = 4 \Delta g_*^\phi / 7$. We find
\be
\Delta N_\nu = \frac{4}{7} g_*(T_r) \left( \frac{g_{*s}(T)}{g_{*s}(T_r)} \right)^{(4+n)/3} \left(\frac{T}{T_r}\right)^{n} \ .
\ee
This is the general expression for the temperature dependent contribution to the number of additional neutrinos. The temperature $T_r$ cannot be much lower than $T_{\rm BBN} \simeq 1 \, {\rm MeV}$, therefore this contribution is vanishing at the time of CMB formation. If we consider $T_r$ around the BBN time, the expression takes the simpler form
\be
\Delta N_\nu \simeq  \frac{4}{7} \frac{43}{4}\left(\frac{T}{T_r}\right)^{n}  \simeq 6.14 \left(\frac{T}{T_r}\right)^{n}  \ .
\label{eq:DeltaNu}
\ee

We impose the recent bound on $\Delta N_\nu$ from Ref.~\cite{Cyburt:2015mya}, where the authors considered an effective number of relativistic species as in \Eq{eq:geffdef}, with $N_\nu$ constant over the different temperature range probed by BBN. Our case is different, since we have a temperature dependent $\Delta N_\nu$. As manifestly shown in \Eq{eq:DeltaNu}, such a correction to the number of SM neutrinos increases with the temperature. In order to put the most conservative limits, we evaluate $\Delta N_\nu$ at a time slightly before neutron freeze-out for temperature $T \simeq 1 \, {\rm MeV}$. At such a temperature, neutrons and protons are still in chemical equilibrium in the entire range for the parameters $(T_r, n)$ under consideration, as explicitly computed in App.~\ref{app:BBN}. In this regard, our bounds are very conservative. Ref.~\cite{Cyburt:2015mya} found the range $2.3 \leq N_\nu \leq 3.4$ at $95 \%$ CL ($2 \sigma$). The contribution in \Eq{eq:DeltaNu} is always positive, so the BBN bounds only allow the region in the $(T_r, n)$ where
\be
T_r \gtrsim (15.4)^{1/n} \; {\rm MeV} \ .
\label{eq:BBNbound}
\ee

\section{Dark Matter Freeze-Out}
\label{sec:FO}

In this section we analyze dark matter freeze-out in the cosmological background introduced in Sec.~\ref{sec:background}. The underlying assumption here is that DM particles achieve thermal equilibrium in the early universe. The conditions needed to satisfy these requirement are given in App.~\ref{app:DMthermalization}. A DM particle interacting through a light mediator (i.e. lighter than the TeV scale) and with coupling strength at least as big as weak gauge interactions thermalizes at temperatures above the TeV scale. In this regime, DM is produced through freeze-out. We first present the Boltzmann equation describing the DM number density evolution. All the results presented in this Section are obtained by numerically solving this equation. In order to understand the qualitative features of the solutions we found, the semi-analytical solution presented in App.~\ref{app:analytical} is very useful. In particular, this solution allows us to estimate the freeze-out temperature and understand the relentless behavior of relics. This regime where DM particles keep annihilating until $T \simeq T_r$ is entered for $n \geq 2$ ($n \geq 4$) if DM annihilations are $s$-($p$-)wave processes. We present explicit solutions for the number density as a function of the temperature, and we quantify the enhancement in the cross section we need with respect to the standard calculation. 

Finally, we investigate the relic density dependence on the DM mass. As is well known, the thermal relic density for WIMPs in a standard cosmology depends on the DM mass very weakly (logarithmic, see App.~\ref{app:analytical}). The quantity that sets the final abundance is the annihilation cross section. We find that this is not the case anymore for a fast expanding universe, since there is a new scale, the temperature $T_r$. The relative hierarchy between the DM mass and $T_r$ determines whether freeze-out happens before or after the epoch of $\phi$ domination. The final relic density differ enormously in the two cases, as we  discuss extensively in this Section.

\subsection{Boltzmann Equation}

From now on, we denote $\chi$ the DM particle, and we assume it to be a Majorana fermion. The DM number density is governed by
\be
\dfrac{dn_\chi}{dt}+ 3 H n_\chi = -\langle \sigma v_{\rm rel}  \rangle \left(n_\chi^{2}-n_\chi^{{\rm eq} \,2}\right) \ .
\label{eq:BoltzmannEq}
\ee
Here, $n_\chi^{{\rm eq}}$ and $\langle \sigma v\rangle$ are the equilibrium number density distribution and the thermally averaged cross section, respectively. This is the same as the standard case~\cite{Lee:1977ua,Scherrer:1985zt,Srednicki:1988ce,Gondolo:1990dk}, with one important difference: the Hubble parameter $H$ is different. Assuming $m_\chi \gg T_r$, the energy density at the freeze-out epoch is dominated by $\phi$ and the Hubble parameters in this regime is given in \Eq{eq:HubbleEarlyTimes}.

As usual, it is convenient to write the \Eq{eq:BoltzmannEq} in terms of the comoving number density, $Y_\chi = n_\chi / s$, and to use $x = m_\chi / T$ as the ``time variable'' 
\be
\dfrac{dY_\chi}{dx}=-\dfrac{s \, \langle \sigma v_{\rm rel} \rangle}{H \, x} \left(1 -  \frac{1}{3} \dfrac{\partial \log g_{*s}}{\partial \log x}\right) \left(Y_\chi^{2} - Y_\chi^{{\rm eq} \,2}\right) \ .
\label{eq:BoltzmannEq2}
\ee
The expression for the comoving equilibrium number density for a Maxwell-Boltzmann distribution is
\be
Y_\chi^{{\rm eq}}(x) =\frac{45 \, g_{\chi}}{4 \pi^4 g_{*s}} x^{2} \BesselK{2}{x} \ ,
\label{eq:Ychieq}
\ee
where $g_\chi = 2$ for a Majorana fermion and $\BesselK{2}{x}$ is the modified Bessel function. At late times the comoving $Y_\chi(x)$ reaches a constant value $Y_\chi(\infty)$, since the actual number density only changes because of the expansion. The present DM density is $\rho_\chi(T_0) = m_\chi \, Y_\chi(\infty) \, s(T_0)$, where $T_0$ is the current temperature of the Cosmic Microwave Background (CMB) photons.

We expand the annihilation cross section times the relative velocity in partial waves
\be
\langle \sigma v_{\rm rel} \rangle = \sigma_s +  \sigma_p \,  x^{-1}  + \mathcal{O}(x^{-2}) \ ,
\label{eq:sigmavpartial}
\ee
keeping only the leading $s$- and $p$-wave contributions. We present numerical results for both cases.

\subsection{An Earlier Freeze-Out}

Before looking at the explicit numerical solution, we examine the qualitative features we expect to find in the solutions. First, and not surprisingly, freeze-out happens  earlier than for the case of a radiation background. This is due to the Hubble parameter during the phase of $\phi$ domination, which for a given temperature is always larger than the associated value in a radiation background. A faster Hubble rate makes it harder for the DM to stay in thermal equilibrium, and freeze-out happens at higher temperatures. 

We provide semi-analytical expressions for the freeze-out temperature in Eqs.~\eqref{eq:FO} and \eqref{eq:FO2} for the case of radiation and modified cosmology, respectively. Keeping the DM mass and the annihilation cross section fixed, and focusing for the purpose of this illustration on $s$-wave processes, the freeze-out temperatures are related by
\be
T_{f \, {\rm rad}}^{1/2} e^{- m_\chi / T_{f  \, {\rm rad}}} = T_f^{1/2} e^{- m_\chi / T_f} \left( \frac{T_f}{T_r} \right)^{n/2} \ .
\ee
Here, $T_r$ and $T_{f \, {\rm rad}}$ are the freeze-out temperature in a generic $(T_r, n)$ and the radiation background, respectively. For freeze-out happening during the $\phi$ dominated epoch, $T_f > T_r$, the freeze-out temperature is larger than the one for the case of a radiation background, $T_f > T_{f \, {\rm rad}}$. Even if the numerical difference between the two temperatures is a factor of a few, the consequent modification of the relic density are significant, since freeze-out happens on the exponential tail of the Maxwell-Boltzmann distribution. 

\subsection{Relentless Relics}

We point out here a very peculiar and previously unrecognized behavior of the number density evolution once $n$ gets large. In order to understand the physics underlying this feature, it is useful to start the discussion by reviewing what happens right after freeze-out for a standard radiation background. DM particles depart from thermal equilibrium when the interaction rate, $\Gamma \simeq n_\chi \langle \sigma v_{\rm rel} \rangle$ is of the order of the Hubble rate, $H_{\rm rad} \simeq T^2 / M_{\rm Pl}$. Immediately after freeze-out, DM particles can still annihilate occasionally, just not enough to stay in thermal equilibrium. The post freeze-out annihilation rate scales as $\Gamma \propto T^3 (T^4)$ for $s$-wave ($p$-wave) annihilations, due to the dilution of the DM particles from the expansion of the universe. This is not enough for the annihilation rate to compete with the  Hubble rate, and post freeze-out annihilations do not change the density significantly. This can be observed in our numerical solutions, and it can also be understood analytically (see \Eq{eq:YchisemilateWAVES}). 

We can repeat the same analysis for the set of modified cosmologies considered here. The argument goes along the same lines, with one important difference: the Hubble parameter now scales as $H \propto T^{2 + n/2}$. Thus there is a critical value of $n$ above which the post freeze-out annihilation rate scales with a power of temperature lower than the one for the Hubble rate. For $s$-wave annihilation, this happens for $n \geq 2$. Interestingly, the case $n = 2$ corresponds to motivated theories of quintessence~\cite{Caldwell:1997ii,Sahni:1999gb}. For $p$-wave annihilation, the condition for this to be the case is $n \geq 4$.

What are the consequences of this relative scaling? For $s$-wave annihilating DM and $n \geq 2$ cosmologies, the annihilation rate red-shifts slower than the Hubble rate. The effects of post freeze-out annihilations is then substantial: DM particles keep annihilating, {\it relentlessly} trying to get to the equilibrium thermal distribution; thermal equilibrium, however, is always unaccessible due to the temperature being low enough for the equilibrium number density to be deeply in the exponential tail. The older the age of the universe, the lower the temperature, and the harder it is for DM particles to get to the equilibrium distribution. The process of depletion goes on until temperatures of the order $T_r$, when the expansion is driven by the radiation bath, and the usual scaling applies. 

\subsection{Number Density Evolution}

\begin{figure}
\center\includegraphics[width=0.5 \textwidth]{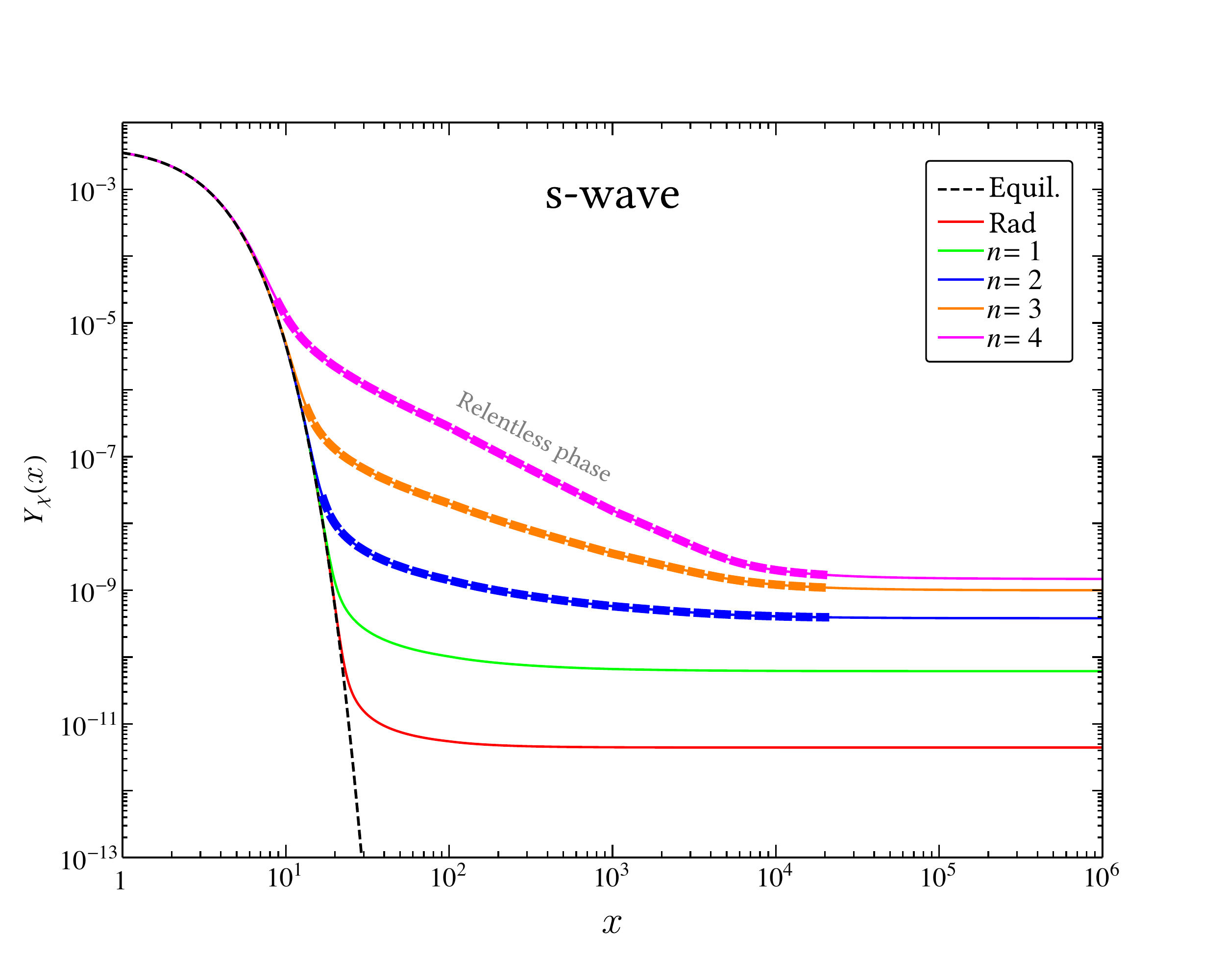}
\caption{Numerical solution of the Boltzmann equation for $m_{\chi}=100 \gev$ and $\langle \sigma v_{\rm rel} \rangle = \sigma_s$. The $s$-wave cross section is fixed to $\sigma_s = 1.7\times10^{-9} \,\gev^{-2}$, which reproduces the observed DM density for a standard cosmology (red line). We set $T_r = 20 \, {\rm MeV}$ for all $n$. We highlight the relentless annihilation phase with thicker dashed lines.}
\label{fig:FreezeOut}
\end{figure}

\begin{figure}
\center\includegraphics[width=0.5 \textwidth]{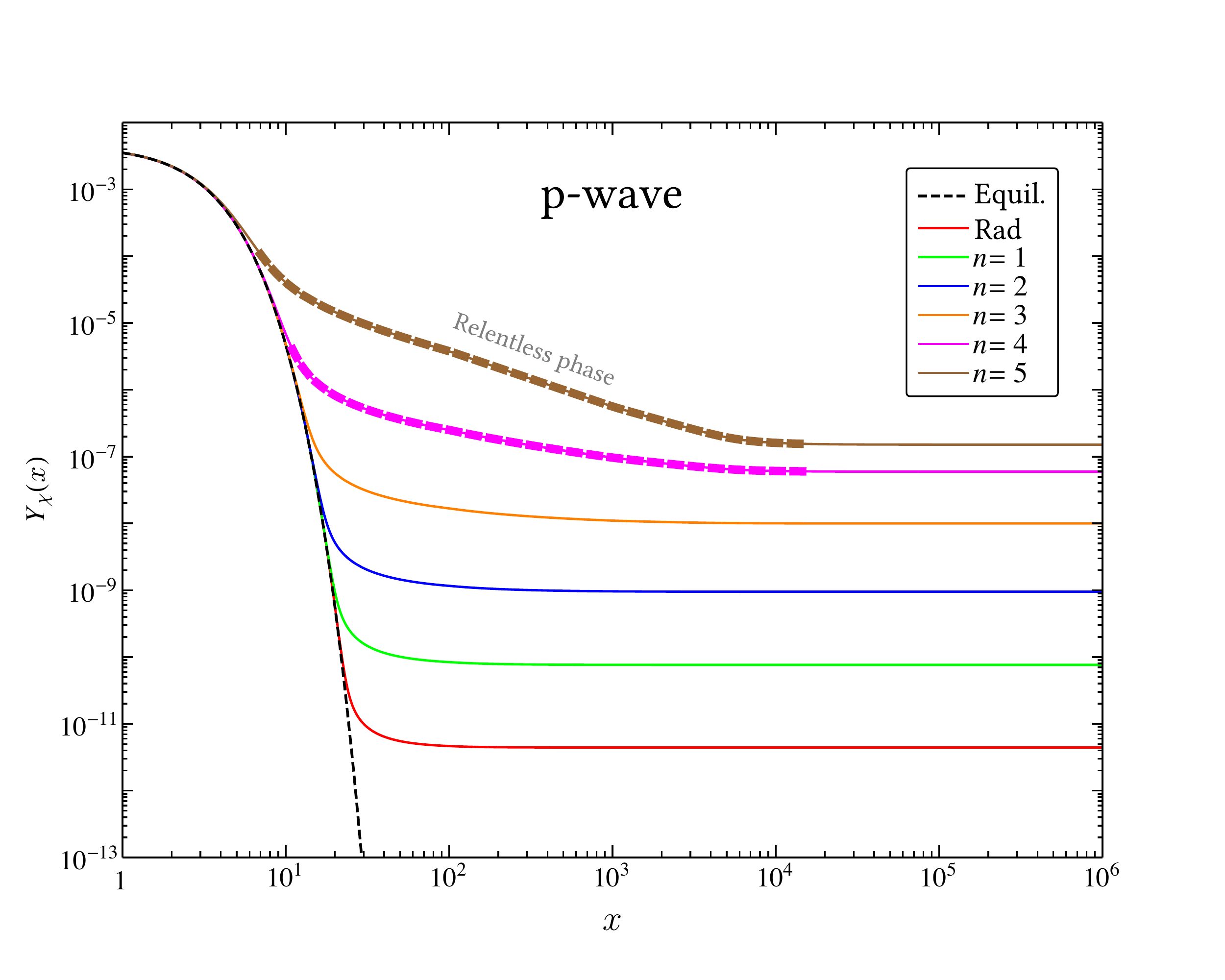}
\caption{Same as Fig.~\ref{fig:FreezeOut} but for $p$-wave annihilation $\langle \sigma v_{\rm rel} \rangle = \sigma_p \, x^{-1}$. The cross section is chosen to reproduce the observed abundance for the standard case, $\sigma_{p}= 7.56\times10^{-8}\gev^{-2}$.}
\label{fig:FreezeOut_p}
\end{figure}

We now show results for the full numerical solutions to the Boltzmann equation in Fig.~\ref{fig:FreezeOut} and \ref{fig:FreezeOut_p} for the case of $s$- and $p$-wave annihilation, respectively. We take a DM mass $m_\chi = 100 \, {\rm GeV}$ (we discuss the very important dependence on mass below) and we fix the annihilation cross section in such a way that we reproduce the observed DM abundance for the case of a standard cosmology (red lines). The solutions for the other cosmological histories are obtained by fixing $T_r = 20 \, {\rm MeV}$ and $n$ as described in the figure caption. 

The comoving number density $Y$ in Fig.~\ref{fig:FreezeOut} do not change significantly after freeze-out for the radiation (red line) and $n = 1$ (green line) cases. This is expected and consistent with the qualitative analysis above. However, an important difference is already clearly visible: freeze-out happens {\em earlier} for $n=1$, than for the $n=0$ standard case, and as a consequence the asymptotic comoving density is higher. 

The phenomenon of {\it relentless} annihilation is visible in Fig.~\ref{fig:FreezeOut} already for the $n = 2$ (thick dashed blue line), as also expected from the discussion above: DM particles continue to find each other to annihilate much later than freeze-out, since the Hubble rate and the red-shifting annihilation rate feature the same scaling with temperature, until the universe becomes radiation dominated and eventually $H\gg\Gamma$. The number density evolution in this regime can be understood analytically (see Eqs.~\eqref{eq:Ychisemilate2} and \eqref{eq:JEFCs}), and it is closely approximated by the expression
\be
Y_{\chi}(x) \simeq \frac{x_r}{m_\chi \, M_{\rm Pl} \, \sigma_s} \left[ \frac{2}{x_f} + \log(x / x_f) \right]^{-1}  \ .
\ee
Here, $x_f$ and $x_r$ are the freeze-out temperature and $T_r$ expressed in terms of the dimensionless variable $x = m_\chi / T$, respectively. The slow logarithmic decrease of the number density is the result of the {\it relentless} attempt of the DM to go back to thermal equilibrium. This behavior persists until $T \simeq T_r$, after which the DM comoving number density reaches a constant value. 

This post freeze-out annihilation are even more pronounced for $n > 2$, as we can see from the orange and the magenta lines in Fig.~\ref{fig:FreezeOut}. In this regime for $n$, the comoving number density is approximated by the following expression
\be
Y_{\chi}(x) \simeq \frac{x^{n/2}_r}{2 \, m_\chi \, M_{\rm Pl} \, \sigma_s} \left[ x_f^{n/2 - 2} + \frac{x^{n/2 - 1}}{n-2}  \right]^{-1}  \ .
\ee
The decrease of $Y_\chi$ is even faster, with a power law instead of the logarithmic dependence appearing for  the marginal case of $n=2$. As before, the number density keeps decreasing with the behavior described above, until radiation takes over.

The discussion for the $p$-wave solutions in Fig.~\ref{fig:FreezeOut_p} is analogous: Freeze-out happens earlier and earlier for higher and higher $n$, and the resulting number density is correspondingly larger. The only difference is that the transition to {\it relentless} relics sets in at $n = 4$, as correctly estimated above. 

\subsection{Enhancement in the Relic Density}

One of the central results of the number density evolution analysis is that freeze-out abundances are in general larger than in the standard case: The red lines is below all the other ones in Figs.~\ref{fig:FreezeOut} and \ref{fig:FreezeOut_p}, for fixed values of the annihilation cross section. One can turn the argument around, and state that {\em larger cross sections} are therefore needed, with the cosmological setup we consider here, to reproduce the observed DM density. This is quite remarkable, as large cross sections translate into larger couplings and therefore larger signals in DM searches, especially in the context of indirect detection. This thus begs the question: How large can the annihilation cross section be, consistently with BBN bounds? 

The two dimensional parameter space $(T_r, n)$ entirely fixes the cosmological background in the present setup. At large values of $T_r$, larger than the DM mass, the standard freeze-out calculation holds, and there is no enhancement to the cross section. The lower $T_r$, the larger the enhancement; However, we cannot take $T_r$ arbitrarily small, as we have to satisfy the BBN bounds in \Eq{eq:BBNbound}.

\begin{figure}
\center\includegraphics[width=0.5 \textwidth]{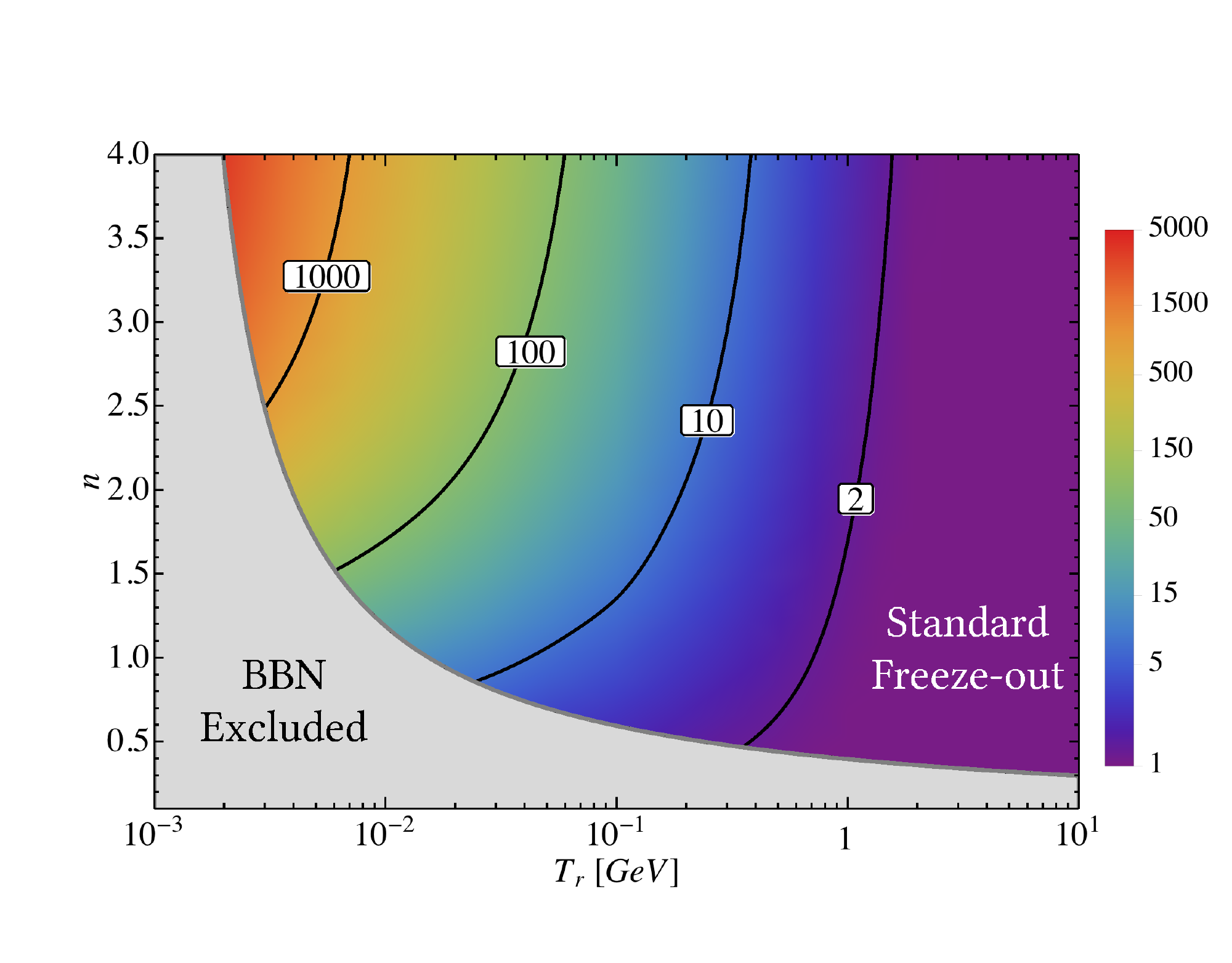}
\caption{Enhancement to the annihilation cross section needed to reproduce the observed DM density due to a cosmological background with a given $T_r$ and $n$. We fix $m_\chi = 100 \, {\rm GeV}$, and we provide the result in units of the $s$-wave cross section $\sigma_s = 1.7\times10^{-9} \,\gev^{-2}$ relative to the standard ($n=0$) radiation-dominated case. The grey region to the bottom left is excluded by BBN.}
%\caption{Enhancement to the thermal relic density due to a cosmological background with a given $T_r$ and $n$, for $m_\chi = 100 \, {\rm GeV}$ and an $s$-wave cross section $\sigma_s = 1.7\times10^{-9} \,\gev^{-2}$ relative to the standard ($n=0$) radiation-dominated case. The grey region to the bottom left is excluded by BBN.}
\label{fig:Cross_Section_s}
\end{figure}

The results for $s$-wave annihilation are shown in Fig.~\ref{fig:Cross_Section_s}, where we fix the DM mass to $m_\chi = 100 \, {\rm GeV}$ and we calculate for each point in the $(T_r, n)$ the cross section needed to produce the right amount of thermal relic DM, normalized to $\sigma_s = 1.7\times10^{-9} \,\gev^{-2}$, the value producing the ``correct'' thermal relic density for a radiation background. We checked numerically that within better than 20\% accuracy, {\em the contour lines also correspond to the enhancement to the thermal relic abundance} for a {\em fixed} value of the pair-annihilation cross section,  in Fig.~\ref{fig:Cross_Section_s} $\sigma_s = 1.7\times10^{-9} \,\gev^{-2}$. In the bottom left corner of the figure we shade in grey the region excluded by BBN.

The figure importantly also indicates the {\em ``boost factors''} expected in indirect detection signals, compared to a standard cosmological setup. %shows contours of constant \sout{cross section needed to get a thermal relic} enhancement of the relic density relative to the standard $n=0$ case. Our results are expressed in units of , therefore the contours in Fig.~\ref{fig:Cross_Section_s} denote the enhancement in the cross section we can afford. 
 The key message is that for the $s$-wave case enhancements beyond $\sim 10^3$ are possible. 

\begin{figure}
\center\includegraphics[width=0.5 \textwidth]{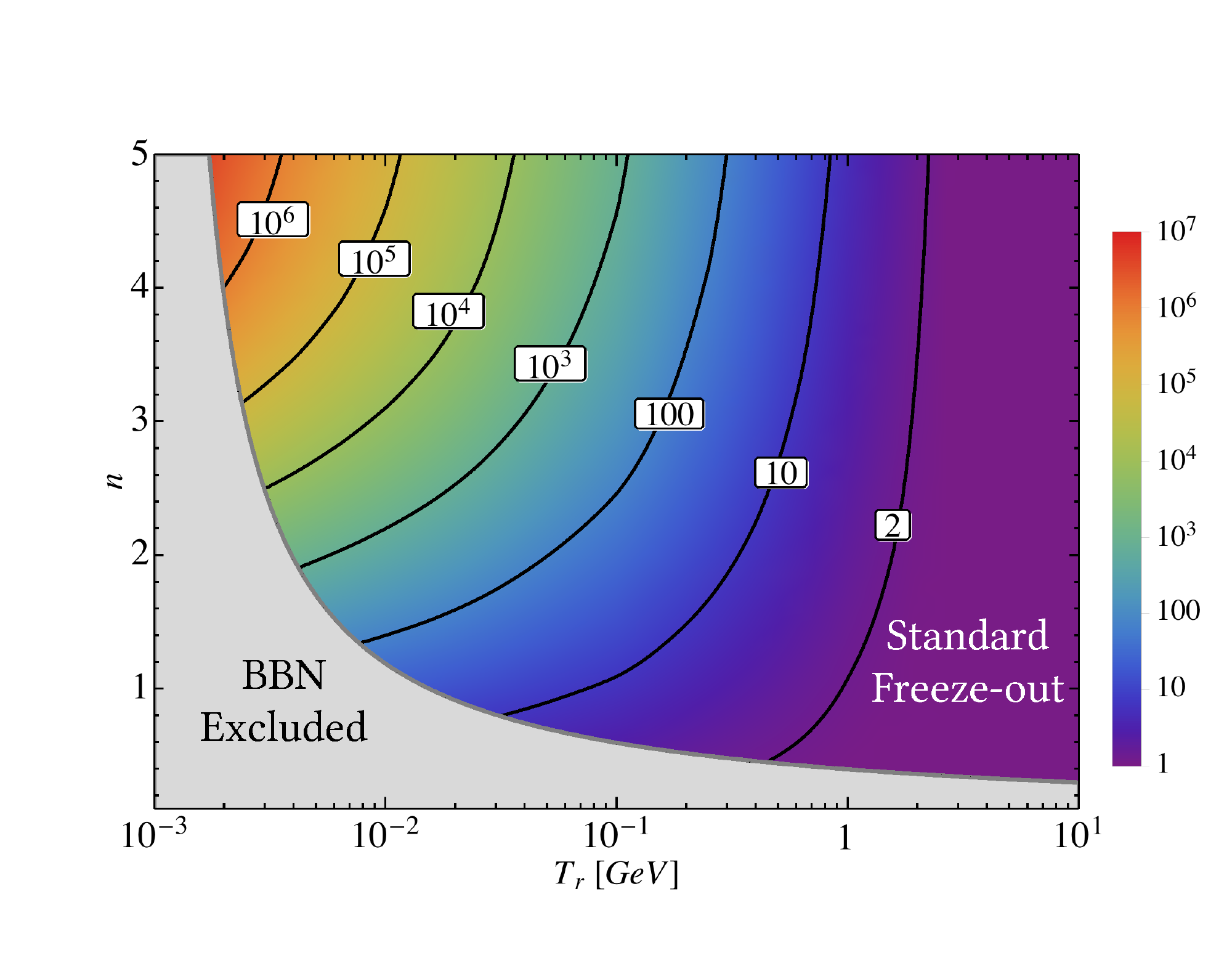}
\caption{As in Fig.~\ref{fig:Cross_Section_s} but for a $p$-wave annihilation cross section that reproduces the observed abundance for the standard case, $\sigma_{p}= 7.56\times10^{-8}\gev^{-2}$.}
\label{fig:Cross_Section_p}
\end{figure}

The analogous analysis for $p$-wave annihilation is presented in Fig.~\ref{fig:Cross_Section_p}. As a result of the temperature dependence of the cross section, {\em larger enhancement factors are possible}, up to $\sim 10^6$ and above. In the case of $p$-wave annihilation, however, indirect signals are suppressed by the DM velocity in the late universe, thus the enhancement to indirect signals is both smaller than the enhancement factors to the relic density, and dependent on environment.

As a side note, we point out that the effect of relentless annihilation produces significantly {\em smaller enhancements than what previously calculated in the literature} (see e.g. \cite{Salati:2002md,Profumo:2003hq}). This is presumably traced back to the previous calculations having assumed a constant value for the comoving number density after a certain effective freeze-out temperature, thus neglecting the {\em relentless} annihilation potentially affecting the relic density over several decades in temperature.

\subsection{Dependence on the DM mass}

All the results presented so far assumed the fixed DM mass benchmark value $m_\chi = 100 \, {\rm GeV}$. For a cold relic in a standard cosmology, the value of the DM mass has a weak impact on the final abundance, which is controlled by the annihilation cross section. We conclude this Section by pointing out one more interesting feature than the framework discussed in this work: the relic density has a strong dependence on the DM mass. 

The reason why this is the case is the presence of the critical temperature $T_r$. If the freeze-out temperature is below $T_r$, there is no change with respect to the standard story. In the opposite case, the precise value of the DM mass is important. Freeze-out happens at temperatures $T_f \simeq m_\chi / 10$, thus the larger the DM mass, the longer the DM particle {\it relentlessly} reduce its comoving number density through residual annihilations. Again, this means that compared to previous calculations the larger the ratio of $m_\chi/T_r$, the larger the effect and the larger the suppression of the calculated enhancement to the thermal relic density.

\begin{figure}
\center\includegraphics[width=0.5 \textwidth]{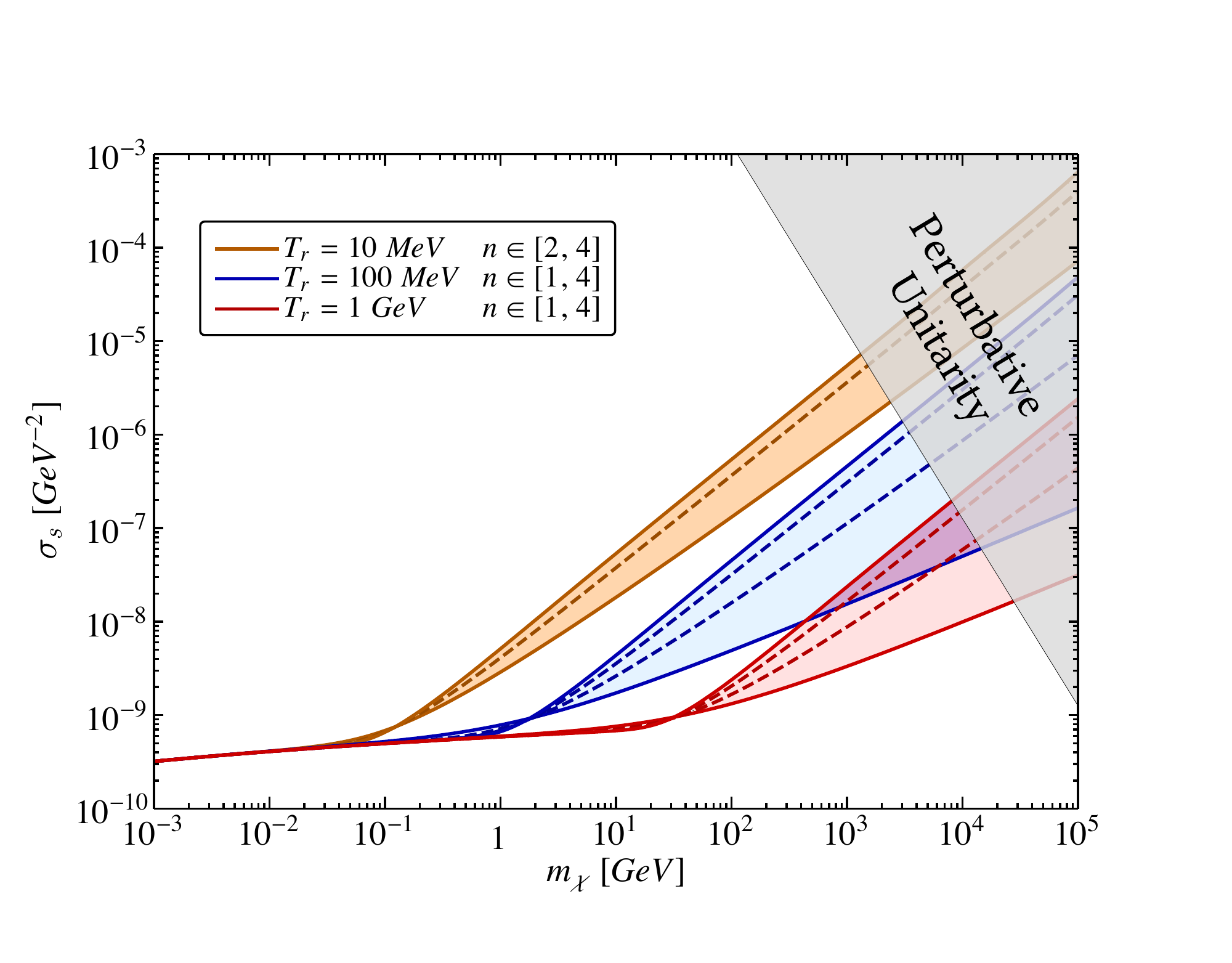}
\caption{Contours of fixed thermal relic density on the $(m_\chi, \sigma_{s})$ plane, for given choices of $n$ and $T_r$, for $s$-wave annihilation cross section (the dashed red and blue lines correspond to, from bottom to top, $n=2$ and 3, while the dashed orange line to $n=3$) . The top right corner is in conflict with limits from perturbative unitarity~\cite{Griest:1989wd}.}
\label{fig:Mass_CrossSection_s}
\end{figure}

\begin{figure}
\center\includegraphics[width=0.5 \textwidth]{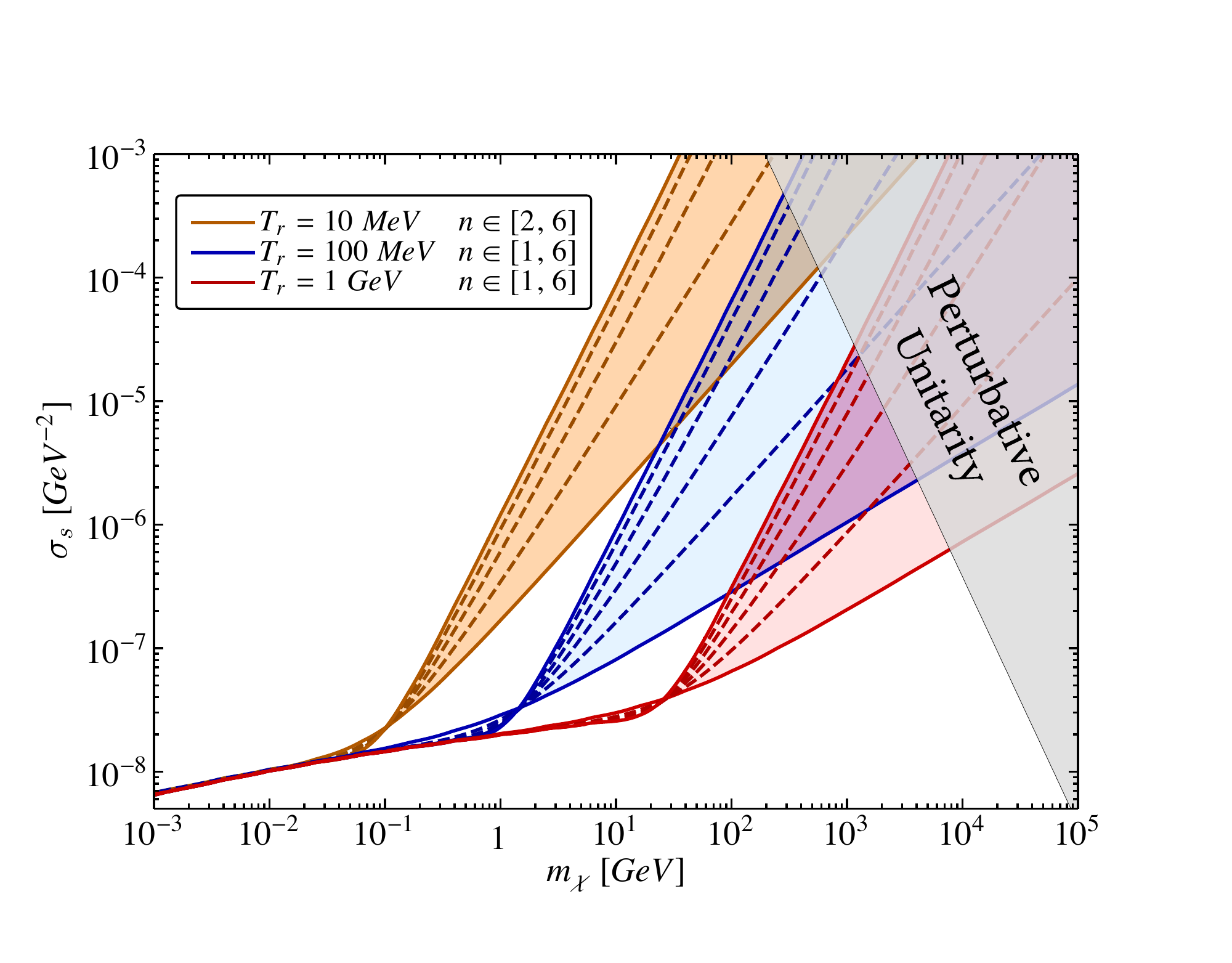}
\caption{As in Fig.~\ref{fig:Mass_CrossSection_s}, but for $p$-wave annihilation cross sections. The intermediate dashed lines, from bottom to top within each shaded region, correspond to increasing integer values of $n$.}
\label{fig:Mass_CrossSection_p}
\end{figure}

To quantitatively study this effect, we fix a few benchmark cosmologies and show contours of fixed relic density in the $(m_\chi, \sigma_{s,p})$ plane. The results are shown in Figs.~\ref{fig:Mass_CrossSection_s} and \ref{fig:Mass_CrossSection_p} for $s$-wave and $p$-wave annihilation cross sections, respectively. At low values of the DM mass, corresponding to a freeze-out temperature below $T_r$, these lines are close to horizontal: This is expected, as in the standard case the relic density depends only on the cross section. The mild dependence on the mass comes from two factors: (i) the logarithmic mass dependence of the freeze-out temperature, and (ii) the different value of $g_*$ at the freeze-out. However, for larger DM mass we see that the relic density strongly depends on the mass, since the larger the DM mass, the longer the phase of  {\it relentless} annihilation, and the ensuing suppression of the relic density. In the figure we also indicate, in the top-right corners, regions in conflict with perturbative unitarity~\cite{Griest:1989wd}. Comparing Figs.~\ref{fig:Mass_CrossSection_s} and \ref{fig:Mass_CrossSection_p} one can also appreciate the steeper dependence on mass in the $p$-wave case. This arises because of the steeper dependence of $\Gamma$ on temperature in the $p$-wave case, and is already reflected in the larger enhancements we find, e.g., in Fig.~\ref{fig:Cross_Section_p} compared to Fig.~\ref{fig:Cross_Section_s}. 

\section{Ultra Stiff Fluids}
\label{sec:EFC}

A virtue of the freeze-out analysis performed in the previous Section is its generality and model independence. Any DM particle thermalized in the cosmological background of \Eq{eq:rhototal} is produced through freeze out as described in Sec.~\ref{sec:FO}. The analysis only assumes our knowledge of the two parameters $(T_r, n)$, without the need of specifying any further property of the new species $\phi$. In this last part of the paper, we provide explicit microscopic realizations for $\phi$, reproducing the red-shift behavior in \Eq{eq:phiScaling}.

All the examples we consider are theories of a single real scalar field $\phi$ minimally coupled to gravity
\be
S= \int d^{4}x\, \sqrt{-g} \left(- \frac{1}{2} g^{\mu\nu} \partial_{\mu} \phi \partial_{\nu} \phi -V(\phi) \right) \ .
\label{eq:Sphi}
\ee
For the remaining of this Section, we set $M_{\rm Pl} = 1$. The energy density and pressure for this fluid read
\beq
 p_{\phi}&=&\frac{1}{2} \left( \frac{d \phi}{d t} \right)^2 - V(\phi) \ ,\\
 \rho_{\phi}&=&\frac{1}{2} \left( \frac{d \phi}{d t} \right)^2 + V(\phi) \ ,
\eeq 
leading to the equation of state 
\beq
w_{\phi} = \dfrac{p_{\phi}}{\rho_{\phi}}=\dfrac{\frac{1}{2} \left( \frac{d \phi}{d t} \right)^2 -V(\phi) }{\frac{1}{2}  \left( \frac{d \phi}{d t} \right)^2 +V(\phi)} \ .
\label{eq:wphidef}
\eeq
For such an equation of state, the energy density red-shifts as $\rho_\phi \propto a^{-3 (1 + w_\phi)}$, which allows us to connect 
\be
n = 3 w_\phi - 1 \ ,
\label{eq:nvsw}
\ee 
where $n$ is the index defined through the red-shift behavior in \Eq{eq:phiScaling}. For a positive scalar potential, the allowed values of $w_\phi$ are in the range $(-1, +1)$. Equivalently, the range for $n$ is between $-4$ and $+2$. The highest $n$ is achieved during a kination phase, where the energy density of $\phi$ is mostly kinetic. In order to get values larger than $n = 2$, we need to consider negative scalar potentials. In what follows, we first describe examples of fluids with $n= 2 $ and then we show how the $n > 2$ domain can be accessed.

\subsection{Quintessence ($n = 2$)}

Examples of theories with $n=2$ are quintessence fluids motivated by the discover of the accelerated expansion of the universe~\cite{Caldwell:1997ii,Sahni:1999gb}. The energy density of this type of fluid red-shifts as $\rho_\phi \propto a^{-6}$ in the kination regime, i.e. when the kinetic energy density dominates over the potential energy . One possible scalar potential leading to this behavior is the exponential form~\cite{Ratra:1987rm,Wetterich:1987fm}
\be
V(\phi) = \exp\left[ - \lambda \phi \right] \ . 
\ee
The role of quintessence for neutralino dark matter freeze-out was studied in Refs.~\cite{Salati:2002md,Profumo:2003hq}.  Alternatives to quintessence, still with the same red-shift behavior, are Chaplygin gas \citep{Kamenshchik:2001cp} or a perfect fluid  described by a polytropic equation of state \citep{Chavanis:2014lra}. 

\subsection{Faster than Quintessence ($n > 2$)}

We provide here example theories where $n > 2$. We assume the energy density of the universe to be entirely dominated by $\phi$, with red-shift as in \Eq{eq:phiScaling}. The scale factor vs time relation can be derived from the Friedmann equation
\beq
a(t)= a_i \left( \frac{t}{t_i} \right)^{2/(n+4)}  \ ,
\eeq
where we define $a_i$ to be the value of the scale factor at $t = t_i$. The time derivative of the Hubble parameter
reads
\be
\frac{d H}{d t} = % \frac{1}{a} \frac{d^2 a}{d t^2} - \left( \frac{1}{a} \frac{d a}{d t} \right)^2 =
 - \frac{1}{2} (\rho_\phi + p_\phi) = - \frac{1}{2} \left( \frac{d \phi}{d t} \right)^2 \ .
 \label{eq:dHdt}
\ee
By comparing this expression with the one resulting from direct calculation, $dH / dt = -2/[(n+4)t^{2}]$, we find the time evolution of the field 
\beq
\phi(t)=\phi_i+\frac{2}{\sqrt{n+4}}\ln\left(\frac{t}{t_i}\right) \ .
\label{eq:phivst}
\eeq

We go back to \Eq{eq:wphidef}, and if we assume that $w_\phi = {\rm const}$ we can solve for the scalar potential
\be
V(\phi) = - \frac{1}{2} \left(\frac{n-2}{n+4}\right)  \left( \frac{d \phi}{d t} \right)^2 \ ,
\ee
where we have traded $w_\phi$ with $n$ by using \Eq{eq:nvsw}. The time derivative of the field $\phi$ is related to the one of the Hubble parameter, as shown in \Eq{eq:dHdt}. We know how the Hubble parameter scales with time in this background with $w_\phi = {\rm const}$, therefore we can find an expression for the potential as a function of time. Once this is done, we use \Eq{eq:phivst} to trade the time variable with $\phi$. The output of this procedure is the scalar potential as a function of the field
\beq
V(\phi) = V_i \, e^{-\sqrt{n+4} \, \phi} \ .
\eeq
The overall constant reads
\beq
V_i = - \frac{2 (n-2) }{(n+4)^2\,t_i^2} \, e^{\phi_i \sqrt{n+4}} \ ,
\eeq
and it is negative for $n > 2$. It is straightforward to check that the solution in \Eq{eq:phivst} with the potential above satisfies the equation of motion $\ddot{\phi}+3H\dot{\phi}+dV(\phi)/d\phi=0$, as it should. This class of potentials have been used in the context of ekpyrotic scenario \cite{Khoury:2001wf}.  A dynamical $w_{\phi}  > 1$ can be obtained also with periodic potentials~\cite{Choi:1999xn,Gardner:2004in}.

\subsection{No superluminal propagation}

We conclude this Section with one important comment. The regime $w_\phi > 1$ implies $p_\phi > \rho_\phi$, and there may be concerns about superluminal propagation. However, the speed of sound for a canonical scalar field with action as in \Eq{eq:Sphi} is always $c_{s}^{2}=1$~\cite{ArmendarizPicon:1999rj,Christopherson:2008ry}. Consequently, causality is not violated.
%\beq
% c^{2}_{s}=\dfrac{\mathcal{L}_{x}}{2X\mathcal{L}_{xx}+\mathcal{L}_{x}}.
%\eeq

\section{Conclusions}
\label{sec:Conclusions}

We analyzed DM freeze-out for non-standard cosmological histories which include a faster-than-usual expansion at early times, driven by a new cosmological species $\phi$. We gave a full description of the cosmological backgrounds in Sec.~\ref{sec:background}. We then parameterized the possible cosmological histories by the values of $n$ and $T_r$, i.e., respectively, the index appearing in \Eq{eq:phiScaling} and the temperature when the energy densities of $\phi$ and radiation are equal. Light element abundances exclude part of this two-dimensional parameter space, and this BBN bound is summarized by \Eq{eq:BBNbound} of Sec.~\ref{sec:BBN}.

In calculating the DM density evolution we identified two distinct possibilities: For $n$ not too large, the behavior is quite similar to the one for standard freeze-out, where shortly after chemical decoupling the comoving number density approaches its asymptotic value. For large $n$, however, we found a new domain where post freeze-out annihilation substantially dilute the DM density. This is explained by the different scaling of the Hubble parameter with  temperature, $H \propto T^{2 + n/2}$; we called this possibility {\it relentless} dark matter. The critical values of $n$ dividing the two regimes are $n = 2$ and $n = 4$ for $s$-wave and $p$-wave annihilation, respectively.

A central result of our analysis is that DM particles which freeze out in the cosmological era dominated by the new species $\phi$ must have cross sections way larger than the thermal value in \Eq{eq:sigmavrelthermal} if DM is to be a thermal relic. We plan to study in the future the implications for dark matter searches of such a large annihilation cross section, such as CMB spectral distortion~\cite{Slatyer:2015jla} and bounds from gamma rays~\cite{Ackermann:2015zua}.

The underlying assumption of our study of DM genesis was an early time thermalization. As discussed in App.~\ref{app:DMthermalization}, this is not necessarily the case, and the faster expansion makes things even harder. If our assumption is not satisfied, DM production would be non-thermal. Assuming production from a decay and/or scattering of particles in the thermal bath, the comoving density produced at a given temperature $T$ approximately reads
\be
Y_\chi(T) \simeq \Gamma(T) H(T)^{-1} \simeq \Gamma(T) \, M_{\rm Pl} \, T^{- 2 - n/2} \ .
\ee
If the rate is mediated by a higher dimensional operator of mass dimension $d$, it would scale as $\Gamma(T) \propto T^{2 d - 7}$. Thus the comoving density at a given temperature scales as $Y_\chi(T) \propto T^{2 d - n/2 - 9}$. We see that the relative size of $d$ and $n$ establishes where most of DM particles are produced. If the dimension is not too large, $d < n/4 + 4.5$, the production is dominated at lower temperature, of the size of the decaying/scattering bath particles. This type of ``IR production'' is known as freeze-in~\cite{Hall:2009bx}. In the opposite case, $d > n/4 + 4.5$, the production is dominated by scattering at high temperatures, similarly to the UV production of axinos or gravitinos~\cite{Rychkov:2007uq,Strumia:2010aa}. This latter case is especially interesting, because it requires the knowledge of how the universe entered the $\phi$ domination phase after inflation. We will study both possibilities in a forthcoming analysis. 

%%%%%%%%

%%%%%%%%%%%%%%%%
\medskip
%%%%%%%%%%%%%%%%%%%%%%%%%%%%%%

\section*{Acknowledgments}
%We acknowledge useful conversations with XXXX. 
This work was supported by the U.S. Department of Energy grant number DE-SC0010107. We are grateful to Josu\'e De Santiago and Simone Ferraro for discussions.

\appendix
 
\section{Dark matter thermalization}
\label{app:DMthermalization}

The DM production mechanism depends on whether the DM ever reaches thermal equilibrium at early times. Thermalization is achieved by collisions, therefore a faster expanding universe makes it harder for the DM to thermalize. This is what we investigate in this Appendix, checking whether the interaction rate between DM and the radiation bath was ever larger than the expansion rate at high temperatures. If this was the case, then DM reaches thermal equilibrium and it is produced through thermal freeze-out. In the opposite case, the production mechanism must be non-thermal. 

For temperatures much larger than the DM mass, the scattering rate can be parameterized as follows
\begin{align}
\label{eq:interactionrate} 
\Gamma_{\rm scatt}(T) \simeq & \, n_{\rm DM} \sigma_{\rm scatt} v_{\rm rel} \simeq \\ & \nonumber 
\frac{3}{2} \frac{\zeta(3)}{\pi^2} T^3 \,  \frac{\lambda^4}{32\pi} \frac{T^2}{(T^2 + M_*^2)^2} \ .
\end{align}
Here, we use the number density for a Majorana fermion in the relativistic regime, and the scattering is assumed to be mediated by a particle with mass $M_*$ that couples to DM and radiation with strength $\lambda$. In what follows, we explore two different possibilities for $M_*$.

\subsection{Massless Mediator}

The first case we study is a massless mediator, $M_* = 0$. Strictly speaking, this analysis is valid also for the case of a massive mediator with mass smaller than the temperatures under consideration. For example, any mediator lighter than the DM particle would fall within this category. The scattering rate in this case reads
\be
\Gamma_{\rm scatt}(T) \simeq \frac{3 \, \lambda^4 \, \zeta(3)}{64 \pi^3} \, T \ , \qquad \quad (M_* \ll T) \ .
\label{eq:interactionrateMassless}
\ee
This linear scaling with the temperature has to be contrasted with the Hubble rate proportional to $T^{2 + n/2}$ (with $n>0$, see Eq.~\eqref{eq:HubbleEarlyTimes}). At high enough temperatures the expansion rate wins, and interactions become more effective as the universe expands and cools down. 

\begin{figure}[!]
\includegraphics[width=0.495 \textwidth]{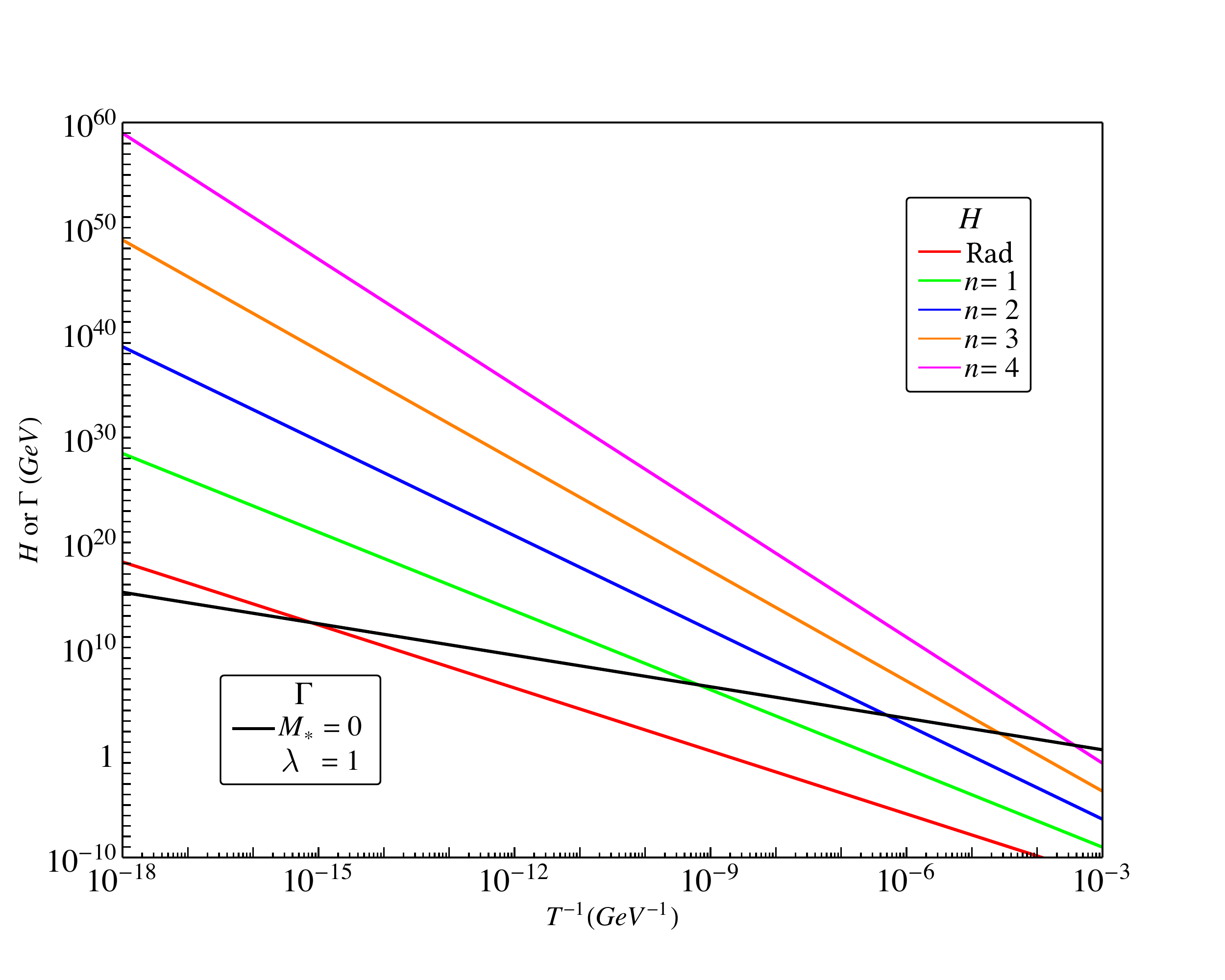} 
\caption{Expansion rate (colored lines) and DM scattering rate (black line) as a function of the inverse temperature (time from left to right). We set the equality temperature between $\phi$ and radiation $T_r = 20 \, {\rm MeV}$, and the coupling strength of the massless mediator $\lambda =1$. DM is in thermal equilibrium for temperatures below the intersection between the black line and the colored line under consideration.}
\label{fig:DMthermalization1}
\end{figure}

A comparison between the Hubble rate in \Eq{eq:HubbleEarlyTimes} for different values of $n$ and the scattering rate in \Eq{eq:interactionrateMassless} is shown in Fig.~\ref{fig:DMthermalization1}, where we plot both these quantities as a function of the inverse temperature. The Hubble rate is obtained by fixing $T_r = 20 \; {\rm MeV}$ in order to have the faster expanding phase to last as long as possible, but still consistent with the BBN bounds discussed in Sec.~\ref{sec:BBN}. The red line corresponds to the standard cosmological history, the other colored line represent the faster expansion rate, with $n$ the index appearing in the exponent of Eq.~\eqref{eq:HubbleEarlyTimes}. The rate is computed by setting the size of the coupling $\lambda \simeq 1$. DM thermalizes at a temperature $T_{\rm th}$ defined to satisfy the condition $H(T_{\rm th}) = \Gamma_{\rm scatt}(T_{\rm th})$. In other words, this temperature can be obtained by finding the intersection between the black lined and the colored line under consideration in Fig.~\ref{fig:DMthermalization1}. This value depends on $n$, and it falls within the range $T_{\rm th} \simeq (10^3, 10^9) \, {\rm GeV}$ as we vary $n$ from $1$ to $4$. DM particles always achieve thermal equilibrium at temperatures higher than the weak scale, even in the extreme case $n = 4$.

\begin{figure}[!]
\includegraphics[width=0.495 \textwidth]{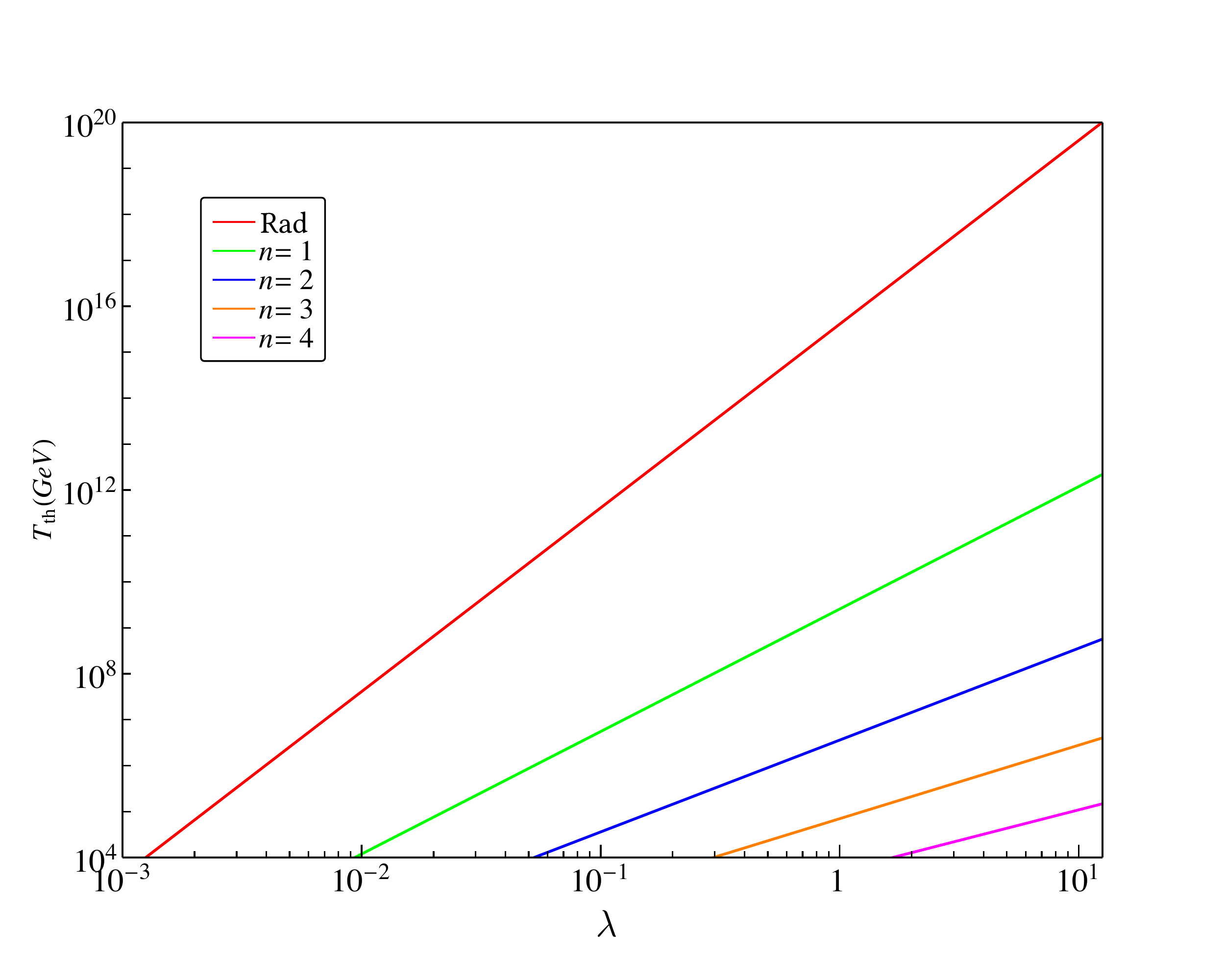}
\caption{DM thermalization temperature for a massless mediator as a function of $\lambda$. Parameters are chosen as in Fig.~\ref{fig:DMthermalization1}.}
\label{fig:DMthermalization2}
\end{figure}

The above conclusion would be altered if we considered smaller values for the coupling $\lambda$. We find it useful to write down an analytical expression for $T_{\rm th}$, which can be obtained by using the approximate expression for the Hubble rate in Eq.~\eqref{eq:HubbleEarlyTimes}. The thermalization temperature approximately reads
\be
T_{\rm th} \simeq \left(  \frac{9 \sqrt{10}  \, \zeta(3)\, \lambda^4}{64 \pi^4 \, \overline{g}^{1/2}_*} \, M_{\rm Pl} \, T_r^{n/ 2}\right)^{2 / (n+2)}   \ .
\label{eq:Tthcase1}
\ee
It scales as $\lambda^{8 / (n+2)}$, so taking a smaller $\lambda$ would affect less the cases of larger $n$. By taking a weak interaction coupling $\lambda \simeq 0.3$, the thermalization temperature is in the range $T_{\rm th} \simeq (10^3, 10^8) \, {\rm GeV}$. The numerical solution for the thermalization temperature as a function of $\lambda$ is shown in Fig.~\ref{fig:DMthermalization2}. For small couplings, $\lambda \lesssim 10^{-3}$, the thermalization temperature is below the weak scale. For weak scale DM this implies that thermal equilibrium is never achieved, and the production mechanism must necessarily be non-thermal.

\subsection{Heavy Mediator}

\begin{figure}
\includegraphics[width=0.495 \textwidth]{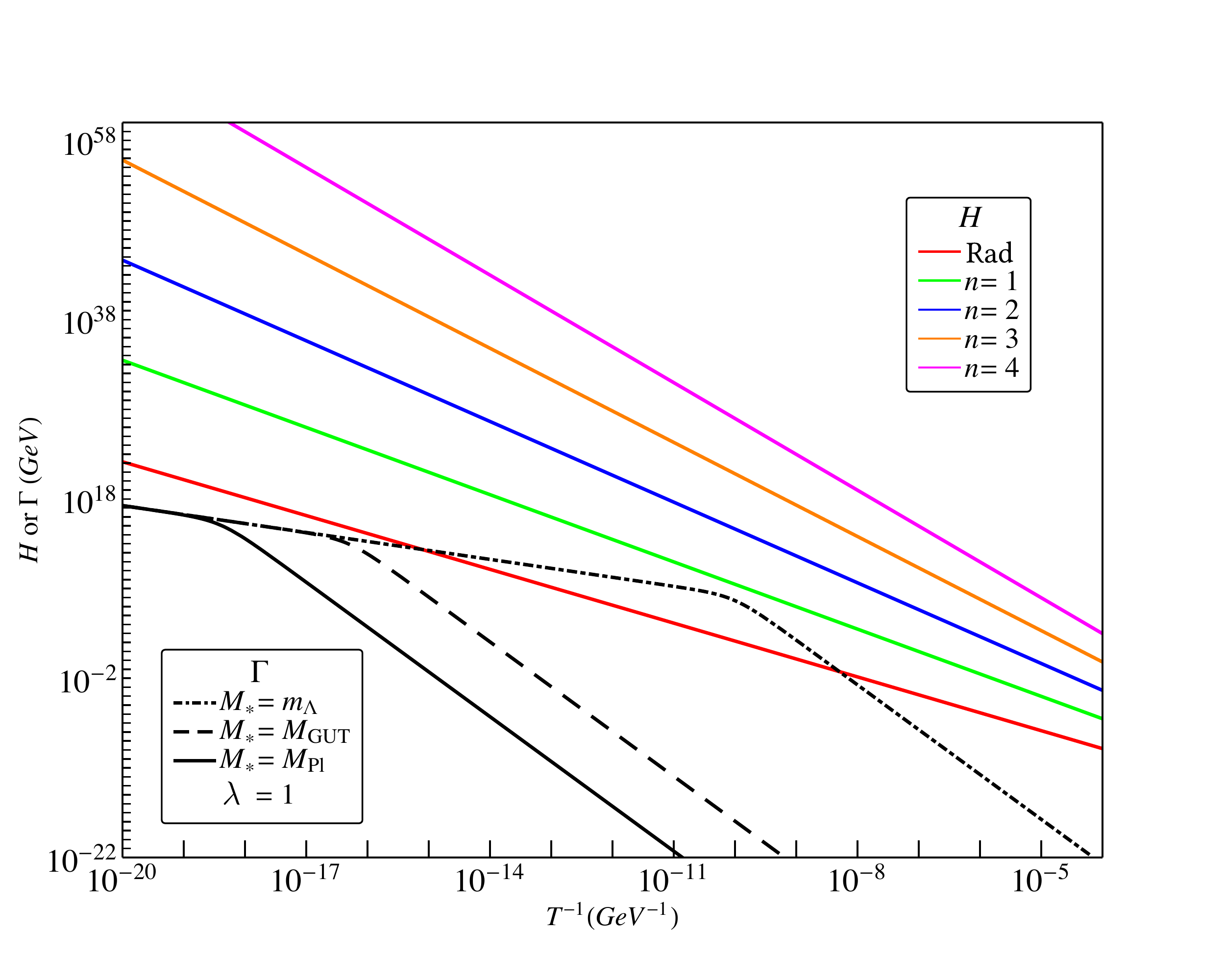}
\caption{Comparison between expansion and interaction rates. Parameters are chosen as in Fig.~\ref{fig:DMthermalization1}, with the only difference that the mediator is massive.}
\label{fig:DMthermalization3}
\end{figure}

We consider here the case of heavy mediators. At temperatures below $M_*$, the scattering rate in approximately
\be
\Gamma_{\rm scatt}(T) \simeq \frac{3 \, \lambda^4 \, \zeta(3)}{64 \pi^3} \, \frac{T^5}{M_*^4} \ , \qquad (T \ll M_*) \ .
\label{eq:interactionrateMassive}
\ee
Unlike the case discussed above, the interaction rate now scales with a higher temperature power than the Hubble parameter. This means that at very early times interactions are effective, and as the temperature drops the expansion takes over. We illustrate this case in Fig.~\ref{fig:DMthermalization3}, where we compare again the rates as a function of the (inverse) temperature. We consider three masses for the mediator. We define $m_\Lambda \simeq 10^{10} \, {\rm GeV}$, the scale where the SM Higgs quartic vanishes~\cite{Buttazzo:2013uya}, hinting for possible new physics~\cite{Hall:2013eko,Hall:2014vga,Fox:2014moa,D'Eramo:2015ssa}. We also consider the unification scale for the gauge couplings ($M_{\rm GUT} \simeq 2 \times 10^{16} \, {\rm GeV}$) and the reduced Planck mass ($M_{\rm Pl})$. For an order one coupling, $\lambda \simeq 1$, thermalization is never achieved for $n>0$. This conclusion is unchanged even if we badly break perturbation theory, $\lambda \simeq 4 \pi$, and it is only strengthened if we consider smaller couplings. We conclude that for a heavy mediator, as heavy as at least $10^{10} \, {\rm GeV}$, DM never equilibrates with the thermal plasma. 

\section{Neutron Freeze-Out and BBN}
\label{app:BBN}

The neutron freeze-out temperature for the cosmological background studied in this work can be found by using the analytical results of Ref.~\cite{Mukhanov:2003xs}. The neutron abundance in conveniently expressed in terms of the neutron fraction $X_n \equiv n_n / (n_n + n_p)$, where $n_{n(p)}$ is the neutron (proton) abundance. The time evolution is described by the Boltzmann equation
\be
\frac{d X_n}{d t} = - \lambda_{n \rightarrow p} \left( 1 + e^{-Q/T} \right) \left( X_n - X_n^{\rm eq} \right) \ .
\ee
Here, we introduce the neutron-proton mass difference
\be
Q \equiv m_n - m_p = 1.293 \; {\rm MeV} \ ,
\ee
and the equilibrium neutron fraction reads
\be
X_n^{\rm eq} = \frac{1}{ 1 + e^{Q/T}} \ .
\ee
For temperatures above the electrons mass, the neutron to proton conversion rates can be approximated by the analytical expression
\be
\lambda_{n \rightarrow p} \simeq 2 \times 1.63 \left( \frac{T}{Q} \right)^3 \left(  \frac{T}{Q} + 0.25 \right)^2 \, {\rm sec}^{-1} \ .
\ee
This simple expression reproduces the full result within the accuracy of a few percent.

The Boltzmann equation for the neutron fraction can be solved as an asymptotic series
\be
X_n = X^{\rm eq}_n \left[ 1 - \frac{\left( 1 + e^{-Q/T} \right)^{-1}}{\lambda_{n \rightarrow p} } \frac{d X^{\rm eq} / dt}{X^{\rm eq}} + \ldots  \right] \ .
\ee
This expression is valid as long as the second term in the parenthesis is smaller than the first, namely if we are close to the equilibrium value. We define the neutron freeze-out as the temperature when the two are equal
\be
\left| \frac{\left( 1 + e^{-Q/T} \right)^{-1}}{\lambda_{n \rightarrow p}(T)} \frac{d X^{\rm eq} / dt}{X^{\rm eq}}\right|_{T = T_n^{{\rm FO}}} \simeq 1 \ .
\ee
The only missing information to solve this equation is the time vs temperature relation, which differs from the one for a standard cosmology due to the presence of $\phi$. This can be derived by imposing conservation of the total entropy. Since we focus on temperatures around the ${\rm MeV}$ scale, we neglect the $g_{*s}$ temperature dependence and the final equation for the freeze-out temperature reads
\be
\left| \frac{\left( 1 + e^{-Q/T} \right)^{-2}}{\lambda_{n \rightarrow p}(T)} 
 \frac{Q}{T} \, H(T) \right|_{T = T_n^{{\rm FO}}} \simeq 1 \ .
 \label{eq:neutronFO}
\ee

The Hubble rate $H$ as defined in \Eq{eq:HubbleRate} contains both the energy density of $\phi$ and radiation. If we only account for radiation and we solve \Eq{eq:neutronFO} we find $T_n^{{\rm FO}} \simeq 0.76 \, {\rm MeV}$, valid for a standard cosmology~\cite{Mukhanov:2003xs}. If we also account for the $\phi$ energy density, we find that this result is not changed by much as long as we consider $T_r \geq 1 \, {\rm MeV}$. More specifically, in the extreme case $n = 0$ (where there is no temperature dependence in $N_\nu$) and $T_r = 1 \, {\rm MeV}$ we find $T_n^{{\rm FO}} \simeq 0.83 \, {\rm MeV}$. In order to put the most conservative bounds, we evaluate $\Delta N_\nu$ as defined in \Eq{eq:DeltaNu} for $T = 1 \, {\rm MeV}$.

\section{Semi-Analytical Freeze-Out}
\label{app:analytical}

In this Appendix we derive semi-analytical solutions for freeze-out relic density. We start with a review of the standard calculation for DM production in a background of radiation, then we extend it to the modified cosmological histories considered in this work. 

\subsection{Standard Cosmology}

In order to connect with the new freeze-out scenarios studied in this paper, we review the Lee-Weinberg calculation for cold relics~\cite{Lee:1977ua}. We simplify the Boltzmann equation in \Eq{eq:BoltzmannEq2} by setting $g_{*} = g_{*s} = {\rm const}$. Furthermore, we Taylor expand the equilibrium density in \Eq{eq:Ychieq} for temperatures lower than the DM mass
\be
Y_\chi^{{\rm eq}}(x) = \frac{45}{4 \, \sqrt{2} \, \pi^{7/2}} \frac{g_\chi}{g_{*}} \, x^{3/2} e^{-x} + \ldots   \qquad (x \gg 1) \ .
\label{eq:YchieqNR}
\ee
The Boltzmann equation can be written as follows
\be
\dfrac{dY_\chi}{dx}= - \frac{}{} 
A \dfrac{\langle \sigma v_{\rm rel} \rangle}{x^2}  \left(Y_\chi^{2} - Y_\chi^{{\rm eq} \,2}\right) \ ,
\label{eq:BEsimple}
\ee
where we define the constant~\footnote{$H_{\rm rad}(x)$ is Hubble parameter obtained by plugging only the energy density of the radiation bath. This is obviously the case for standard freeze-out. We find this definition useful also for the case when the energy density is dominated by $\phi$.}
\be
A = \frac{s(m_\chi) }{H_{\rm rad} (m_\chi) }  = 
\frac{2 \sqrt{2} \, \pi}{3 \sqrt{5}} \, g_*^{1/2} \, m_\chi M_{\rm Pl}  \ .
\label{eq:defA}
\ee

We identify two distinct regimes for the solution. At early times, DM annihilations are efficient and $Y_\chi$ closely tracks the equilibrium distribution. At late times, the expansion takes over and the density freezes-out. We solve the Boltzmann equation in these two regimes and then match the two solutions at some intermediate point. We perform the matching at the temperature where $Y_\chi$ moves away from its equilibrium expression, a point known as the DM freeze-out. 

We find it convenient to write the Boltzmann equation for $\Delta_\chi \equiv Y_\chi - Y_\chi^{\rm eq}$, which is obtained by plugging its definition into \Eq{eq:BEsimple}. We find 
\be
\frac{d \Delta_\chi}{d x} = - A \frac{\langle \sigma v_{\rm rel} \rangle}{x^2} \, \Delta_\chi (2 Y_\chi^{\rm eq} + \Delta_\chi) - \frac{d Y_\chi^{\rm eq} }{d x}  \ .
 \label{eq:BEsimple2}
\ee
At times much earlier than freeze-out, the departure from thermal equilibrium is minimal and we can neglect terms quadratic in $\Delta_\chi$ and its derivative. As a consequence, the DM number density can be approximated by
\be
Y_\chi(x) \simeq Y_\chi^{\rm eq}(x) + \frac{x^2}{2 A \langle \sigma v_{\rm rel} \rangle} \qquad  (1 < x < x_f) \ .
\label{eq:Ychisemiearly}
\ee
In the opposite regime, we neglect the equilibrium distribution in the Boltzmann equation~\eqref{eq:BEsimple}, which can be integrated to find the solution
\be
Y_\chi(x) \simeq \left[ \frac{1}{Y_\chi(x_f)} + A \, J(x) \right]^{-1} \qquad  (x > x_f) \ .
\label{eq:Ychisemilate}
\ee
Here, we define the annihilation integral
\be
J(x) \equiv \int_{x_f}^x \frac{\langle \sigma v_{\rm rel} \rangle}{x^2} dx  \ .
\ee

The term equal proportional to the inverse comoving density at the freeze-out in \Eq{eq:Ychisemilate} is important to ensure that our solution is continuous. However, it is numerically subdominant, unless we consider values $x \simeq x_f$. This can be explicitly checked for the partial wave expansion of \Eq{eq:sigmavpartial}, for which the annihilation integral now reads:
\be
J(x) \simeq \sigma_s \left( \frac{1}{x_f} - \frac{1}{x} \right) + \frac{\sigma_p}{2}   \left( \frac{1}{x^2_f} - \frac{1}{x^2} \right) \ . 
\ee
The comoving number density after freeze-out reads
\be
Y_\chi(x) = \frac{x_f}{A} \left\{ \begin{array}{cccc}
\frac{\left( 1 - x_f / x \right)^{-1}}{\sigma_s} & & & \text{s-wave} \\
\frac{2 x_f \, \left( 1 - (x_f / x)^2 \right)^{-1}}{\sigma_p} & & & \text{p-wave}
\end{array} \right. \ .
\label{eq:YchisemilateWAVES}
\ee
The above equation illustrates how the comoving number density quickly approaches a constant values after freeze-out. This is only valid for the standard case of a radiation background. In the cosmological histories discussed in this work, we find that DM particles keep annihilating well after the number density has departed from its equilibrium value.

The current DM abundance is evaluated from the asymptotic value ($x \gg x_f$) of the comoving number density. This can be obtained by extrapolating \Eq{eq:Ychisemilate} to very large values of $x$, and we find
\be
Y_\chi^\infty = \frac{1}{A \, J(\infty)} = \frac{3 \sqrt{5}}{2 \sqrt{2} \, \pi} \,\frac{\left( \frac{\sigma_s}{x_f} + \frac{\sigma_p}{2 x^2_f} \right)^{-1}}{g_*^{1/2} \, m_\chi M_{\rm Pl}} \ .
\ee
The first equality is general, whereas the second assumes the solutions in \Eq{eq:YchisemilateWAVES} for a partial wave expansion. The asymptotic number density scales as the inverse DM mass. In units of the critical density, the DM density results in
\be
\Omega_\chi h^2 \equiv \frac{m_\chi Y_\chi^\infty \, s(T_0)}{\rho_{\rm cr} / h^2} = 
\frac{2 \times 10^8 \, {\rm GeV}^{-1}}{M_{\rm Pl} \, g_*^{1/2} \,  \left( \frac{\sigma_s}{x_f} + \frac{\sigma_p}{2 x^2_f} \right)} \ .
\ee
This quantity depends on the DM mass only through the value of $x_f$.

Finally, we determine the value of the freeze-out temperature. This is the point where we match the two solutions in Eqs.~\eqref{eq:Ychisemiearly} and \eqref{eq:Ychisemilate}. We define the freeze-out as temperature $x_f$ by imposing $\Delta_\chi(x_f) = c \,Y_\chi^{\rm eq}(x_{f})$, where $c$ is an order one coefficient. We plug this definition into the Boltzmann equation~(\ref{eq:BEsimple2}), and the freeze-out condition is expressed as follows
\be
\left.  \frac{e^x \, x^{1/2}}{\langle \sigma v_{\rm rel} \rangle} \right|_{x = x_f}  =
\frac{c(c+2)}{c+1} \frac{3 \sqrt{5}}{2 \, \pi^{5/2}} \, \frac{g_\chi}{g_*^{1/2}} \, m_\chi \, M_{\rm Pl} \ ,
\label{eq:FO}
\ee
where we also restore the definition for $A$ as in \Eq{eq:defA}. We remind that the thermally averaged cross section can depend on $x$, as in the case of $p$-wave annihilation.

\subsection{Non-Standard Cosmology Freeze-Out}

For the modified cosmological backgrounds considered here, the DM number density evolution is still described by \Eq{eq:BoltzmannEq2}. However, the temperature dependence of the Hubble parameter is different. We introduce the quantity $x_r \equiv m_\chi / T_r$, where $T_r$ was defined as the temperature where the energy of the radiation bath reaches the one of $\phi$. The Boltzmann equation now reads
\be
\dfrac{dY_\chi}{dx}= - \frac{}{} 
A \dfrac{\langle \sigma v_{\rm rel} \rangle}{x^{2 - n/2} \left(x^n + x_r^n \right)^{1/2}}  \left(Y_\chi^{2} - Y_\chi^{{\rm eq} \,2}\right) \ ,
\label{eq:BEsimpleNS}
\ee
where we use again the parameter $A$ defined in \Eq{eq:defA}. 

We assume that freeze-out happens during the time of $\phi$ domination, namely $x_f \ll x_r$. At the freeze-out time, the Boltzmann equation can then be approximated by
\be
\dfrac{dY_\chi}{dx} \simeq - \frac{}{} 
A \dfrac{\langle \sigma v_{\rm rel} \rangle}{x^{2 - n/2} \, x_r^{n/2}}  \left(Y_\chi^{2} - Y_\chi^{{\rm eq} \,2}\right) \ .
\label{eq:BEsimpleNSFO}
\ee
We solve again before and after freeze-out by using the convenient variable $\Delta_\chi$. At earlier times we neglect terms quadratic in $\Delta_\chi$ and its derivative
\be
Y_\chi(x) \simeq Y_\chi^{\rm eq}(x) + \frac{x^{2 - n/2} x_r^{n/2}}{2 A \langle \sigma v_{\rm rel} \rangle}   \qquad  (1 < x < x_f)\ .
\label{eq:Ychisemiearly2}
\ee
After freeze-out, the solution takes the same form 
\be
Y_\chi(x) \simeq \left[ \frac{1}{Y_\chi(x_f)} + A \, J_\phi(x) \right]^{-1} \quad  (x_f < x < x_r) \ .
\label{eq:Ychisemilate2}
\ee
This looks analogous to \Eq{eq:Ychisemilate}, but with the crucial difference that the annihilation integral reads
\be
J_\phi(x) \equiv \frac{1}{x_r^{n/2}} \int_{x_f}^x \frac{\langle \sigma v_{\rm rel} \rangle}{x^{2 - n/2}} dx  \ .
\ee
We can perform the integral for partial wave expansion, and we find the expressions
\be
J^{({\rm s})}_\phi(x) = \frac{\sigma_s}{x_r^{n/2}} \left\{
\begin{array}{cccl}
\frac{x_f^{n/2-1} \, - \, x^{n/2 - 1}}{1 - n /2} & & & n \neq 2 \\
\log(x / x_f) & & & n = 2 
\end{array}
\right. \ ,
\label{eq:JEFCs}
\ee
and 
\be
J^{({\rm p})}_\phi(x) = \frac{\sigma_p}{x_r^{n/2}} \left\{
\begin{array}{cccl}
\frac{x_f^{n/2-2} \, - \, x^{n/2 - 2}}{2 - n /2} & & & n \neq 4 \\
\log(x / x_f) & & & n = 4
\end{array}
\right. \ ,
\ee
for $s$- and $p$-wave, respectively. 

The solution in \Eq{eq:Ychisemilate2} can only be extrapolated up to $x = x_r$. Once the radiation bath dominates the energy density, we perform an additional matching, analogous to the one for standard freeze-out (see \Eq{eq:Ychisemilate}). The subsequent evolution is described by 
\be
Y_\chi(x) \simeq \left[ \frac{1}{Y_\chi(x_r)} + A \, J_{\rm rad}(x) \right]^{-1} \qquad  (x > x_r) \ ,
\label{eq:Ychisemiverylate2}
\ee
where define the annihilation integral now reads
\be
J_{\rm rad}(x) \equiv \int_{x_r}^x \frac{\langle \sigma v_{\rm rel} \rangle}{x^2} dx  \ .
\ee

The final DM density is $\rho_\chi(t_0) = m_\chi Y_\chi^\infty \, s(T_0)$, where the asymptotic value of the comoving density can be extracted by \Eq{eq:Ychisemiverylate2}. 

We conclude with the evaluation of the freeze-out temperature, defined as before by the condition $\Delta_\chi(x_f) = c \,Y_\chi^{\rm eq}(x_{f})$. We find
\be
\left.  \frac{e^x \, x^{1/2}}{\langle \sigma v_{\rm rel} \rangle} \left(\frac{x_r}{x}\right)^{n/2} \right|_{x = x_f}  =
\frac{c(c+2)}{c+1} \frac{3 \sqrt{5}}{2 \, \pi^{5/2}} \, \frac{g_\chi}{g_*^{1/2}} \, m_\chi \, M_{\rm Pl} \ .
\label{eq:FO2}
\ee
This relation is very similar to \Eq{eq:FO2} with the important difference of a $(x_r/x)^{n/2}$ factor, which significantly enhances the left-hand side since we consider freeze-out during the $\phi$ domination phase ($x_f \ll x_r$). If we fix the DM mass and annihilation cross section, freeze-out must happen earlier with respect to the standard case.

\bibliography{EFCpaper1}

\end{document}